\documentclass[aps,reprint,longbibliography]{revtex4-1}
\usepackage{amsmath,amsfonts,amssymb}
\usepackage{graphicx}
\usepackage[unicode=true,colorlinks=true,citecolor=blue,urlcolor=blue]{hyperref}
\usepackage{bm}

\renewcommand{\Re}{\mathop{\rm Re}}
\renewcommand{\Im}{\mathop{\rm Im}}
\newcommand{\Tr}{\mathop{\rm Tr}}
\renewcommand{\sinh}{\mathop{\rm sh}}
\renewcommand{\cosh}{\mathop{\rm ch}}
\renewcommand{\tanh}{\mathop{\rm th}}

\newcommand{\e}{\mathrm{e}}

\renewcommand{\i}{{\rm i}}
\renewcommand{\d}{\mathrm d}

\newcommand{\Js}{{\bm{\mathcal{J}}}}
\newcommand{\Jdr}{{\bm{\mathcal{J}}}_{\rm dr}}
\newcommand{\Jdiff}{{\bm{\mathcal{J}}}_{\rm diff}}

\graphicspath{{img/}}

\newcommand{\nix}[1]{}

\usepackage[normalem]{ulem}

\begin{document}

\title{Spin-related phenomena in two-dimensional hopping regime in magnetic field}

\author{A.\,V.  Shumilin}
\author{D.\,S. Smirnov}
\author{L.\,E. Golub}
\affiliation{Ioffe Institute, 194021 St.~Petersburg, Russia}


\begin{abstract}

The spin relaxation time of localized charge carriers is few orders of magnitude larger than that of free electrons and holes. Therefore mutual conversion of spin polarization, charge current and spin current turns out to be underlined in the hopping conductivity regime. We reveal different regimes of the coupled spin and charge dynamics depending on the relation between spin relaxation time and the characteristic hopping time. We derive kinetic equations to describe electrical spin orientation, dc spin-Hall effect, and spin galvanic effect in the transverse magnetic field. The generalized macroscopic conductivities describing these effects are calculated using percolation theory supported by numerical simulation. The conductivities change the sign at least once as functions of magnetic field for all values of the spin relaxation time.

\end{abstract}


\maketitle{}


\section{Introduction}

\begin{figure}[t]
  \centering
  \includegraphics[width=\linewidth]{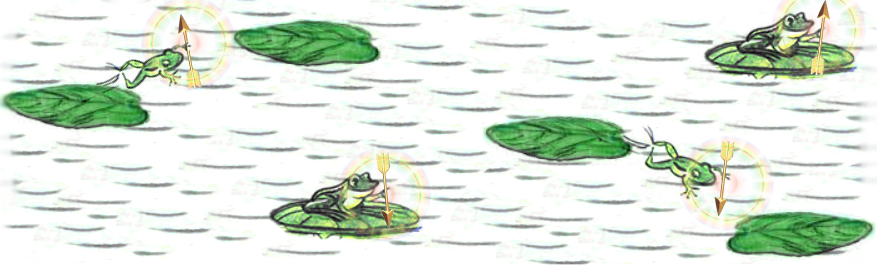}
  \caption{Impression of SHE effect. The hopping of electrons (frogs) in one direction is accompanied by separation of electrons (frogs) with spins (arrows) up and down in perpendicular direction.}
  \label{fig:frogs}
\end{figure}

Spin is in the center of condensed-matter physics for almost two decades due to remarkable effects allowing for both deeper understanding of fundamental physical processes and some possible future applications~\cite{dyakonov_book}. One of the most investigated spin-related phenomena is the spin-Hall effect (SHE) which is a conversion of an electric current into spin current~\cite{dyakonov71,dyakonov71a,kato04,Schliemann2006}.
There is also an inverse effect (inverse SHE) consisting in the generation of the electric current under the spin current flow~\cite{dyakonov71,kkm94,doi:10.1063/1.2199473,PhysRevLett.98.156601}.
The SHE is qualitatively similar to the ordinary Hall effect: The electric current in the system is converted into the spin current or spin-up and spin-down separation in the perpendicular direction. This means that the charge carriers with opposite spins flow preferentially in opposite directions. Impression of this effect is presented in Fig.~\ref{fig:frogs}. Microscopically SHE arises due to spin-orbit interaction, and it is symmetry allowed in any system.
There are some more subtle spin-dependent phenomena which take place only in systems of low point symmetry. {The first example is} the current-induced spin orientation (CISP) consisting in the generation of a net spin polarization by electric {current~\cite{ivchenko1978new,vorob1979optical,vasko1979spin,Levitov1985,aronov1989spin,edelstein90,aronov1991spin}.} The reciprocal phenomenon, the Spin-galvanic effect (SGE), is a generation of electrical current in the process of nonequilibrium spin relaxation~\cite{Ivchenko_Ganichev_in_Dyakonov_book}. Both CISP and SGE are symmetry-allowed in gyrotropic (optically active) systems. They have been investigated in gyrotropic bulk semiconductors, for example, {tellurium~\cite{ivchenko1978new,vorob1979optical}}, strained zinc-blende III-V {crystals~\cite{silov04,Ganichev_110,PhysRevLett.93.176601}} and in various two-dimensional (2D) heterostructures~\cite{Ganichev_Trushin_Schliemann}. CISP and SGE can be viewed as the consequences of SHE (or inverse SHE), so all three spin-related phenomena are {interconnected~\cite{Hopping_spin}}. The microscopic source for the conversion of the spin current into the net spin polarization (CISP) and to electric current (SGE) is the spin-momentum linear coupling caused by Rashba- and 2D Dresselhaus spin-orbit {interactions~\cite{ohkawa1974quantized,Vasko79,dyakonov:110,golub_ganichev_BIASIA}}.

Besides the spin-dependent effects related with the electric current flow, there is a reach spin physics of carriers localized at {neutral dopants,} interfaces of semiconductor heterostructures and in quantum dots. These systems attract permanent interest due to long spin relaxation times which can be by orders of magnitude larger than for free carriers and vary in a broad range~\cite{KKavokin-review}. The reason for long spin lifetimes is that the major mechanisms of spin relaxation related with free carrier momentum scattering are absent for localized carriers, and {spin relaxation is determined by a weak hyperfine interaction with host lattice nuclei~\cite{merkulov02,PhysRevLett.88.186802}.} Long spin memory allows for {fast spin manipulation by optical pulses~\cite{gywat2009spins,glazov:review,warburton2013}, resonant spin amplification~\cite{Kikkawa98,beschoten}, electron spin precession mode locking~\cite{A.Greilich07212006,yugova09,sokolova09,fokina-2010,yugova12}, nuclei induced frequency focusing~\cite{A.Greilich09282007,glazov2010a,carter:167403,2010arXiv1006.5144K,yugova11} and measurement of spin fluctuations~\cite{crooker2010,Zapasskii:13,Oestreich-review,eh_noise}.}

The two groups of the above-described spin-dependent effects, related with the electrical current flow and with the long-lived localized spins, meet in systems with hopping conductivity. Indeed, if the localized carriers can migrate between the localization sites then one can study SHE, CISP and SGE in systems with slow spin relaxation. Recently we have shown that all three effects take place in  2D systems with hopping conductivity and pronounced spin-orbit interaction~\cite{Hopping_spin}. In the present work, we investigate these spin-related phenomena for localized carriers, as functions of the nuclei-induced spin relaxation time and external perpendicular magnetic field.

The paper is organized as follows.
In Sec.~\ref{sec:general} we derive and analyze kinetic equations for the coupled charge and spin dynamics.
In Sec.~\ref{sec:averaging} we solve these equations using both numerical simulation and percolation analysis. The obtained results and their generalizations are discussed in Sec.~\ref{sec:Zeeman} and are  summarized in Sec.~\ref{sec:concl}.

\section{General theory}
\label{sec:general}

\subsection{Phenomenology}
\label{sec:phenomen}

CISP, SGE and SHE are introduced by the following phenomenological expressions~\cite{Hopping_spin}
\begin{equation}
	\bm s = \hat{\bm \sigma}_\text{CISP} \bm E,
	\quad
		\bm j = \hat{\bm \sigma}_\text{SGE} \bm s,
	\quad
		\bm{\mathcal J} = \hat{\bm \sigma}_\text{SHE} \bm E,
\label{eq:3effects}
\end{equation}
{where $\bm s$ is the average spin polarization, $\bm E$ is the applied electric field, $\bm j$ is the current density, and $\bm{\mathcal J}$ is the spin current associated with spin component perpendicular to the structure.} The generalized conductivities $\hat{\bm \sigma}_\text{CISP}$, $\hat{\bm \sigma}_\text{SGE}$ and $\hat{\bm \sigma}_\text{SHE}$ depend on structure parameters and external magnetic field $\bm B$.

We consider a semiconductor zinc-blende heterostructure  grown along the $[001]$ direction. In this case it is useful to introduce the coordinate frame as $z\parallel[001]$, $x \parallel [1\bar{1}0]$ and $y \parallel [110]$. In this coordinate frame the Hamiltonian describing spin-orbit interaction has the form~\cite{golub_ganichev_BIASIA}
\begin{equation}
\label{H_SO}
	{\cal H}_\text{SO} = \bm{\hat{\sigma} \cdot \hat{\beta}\bm k} =
\hat{\sigma}_x \beta_{xy} k_y + \hat{\sigma}_y \beta_{yx} k_x .
\end{equation}
Here $\hat{\sigma}_{x,y}$ are the Pauli matrices, 
$\bm k = -i\bm\nabla$, and {$\beta_{xy}$, $\beta_{yx}$ are  two} spin-orbit constants caused by both bulk- and structure-inversion asymmetry. We assume that the external magnetic field is applied perpendicular to the structure, ${\bm B=B_z \bm e_z}$ where $\bm e_z$ is the unit vector along $z$ direction.

The asymmetric heterostructures are described by C$_{2v}$ point symmetry group. In this case the components $E_x$ and $s_y$ transform according to $\Gamma_2$ representation, $E_y$ and $s_x$ belong to $\Gamma_4$ representation, while $B_z$ transforms according to $\Gamma_3$. {Importantly,} the symmetry analysis shows {that
  \begin{itemize}
  \item the diagonal components of all the generalized susceptibilities $\hat{\bm \sigma}$ are odd in $B_z$,
  \item the off-diagonal components are even in $B_z$.
  \end{itemize}
In} the particular cases of structure inversion asymmetry  dominance (C$_{\infty v}$ symmetry) and bulk inversion asymmetry dominance  (D$_{2d}$ symmetry), the components of the tensors $\hat{\bm \sigma}_\text{CISP,SGE}$ are related by
\begin{equation}
  \sigma_\text{CISP,SGE}^{xy} = \mp \sigma_\text{CISP,SGE}^{yx}, \quad
  \sigma_\text{CISP,SGE}^{xx}= \pm \sigma_\text{CISP,SGE}^{yy},
\end{equation}
where the upper (lower) sign should be taken for C$_{\infty v}$ (D$_{2d}$) point symmetry.
For the spin-Hall effect, the following relation takes place in both cases:
\begin{equation}
  \sigma_\text{SHE}^{xy} = -\sigma_\text{SHE}^{yx}, \quad
  \sigma_\text{SHE}^{xx}= \sigma_\text{SHE}^{yy}.
\end{equation}
The structures grown along {crystallographic directions other than $[001]$} are briefly discussed in Sec.~\ref{sec:Zeeman}.

\subsection{Derivation of the kinetic equation}
\label{sec:micro}

In the hopping conductivity regime the electron energies are different for different localization sites. Therefore hopping {between the sites} involves emission or absorption of phonons to ensure the energy conservation. The total Hamiltonian of the system can be presented as
\begin{equation}
  {\cal H}={\cal H}_e+{\cal H}_{ph}+{\cal H}_{e-ph}.
\end{equation}
Here the term ${\cal H}_e$ describes the Hamiltonian of the electronic system, ${\cal H}_{ph}$ is the phonon Hamiltonian, and ${\cal H}_{e-ph}$ describes the electron-phonon interaction.

The Hamiltonian describing the system of localized electrons reads
\begin{equation}
 {\cal H}_e=\sum_{i,\sigma}\epsilon_ic_{i\sigma}^\dag c_{i\sigma}+\sum_{ij}\sum_{\sigma\sigma'}{J}_{ij}^{\sigma\sigma'}c_{i\sigma}^\dag c_{j\sigma'}+\mathcal H_Z.
\label{eq:ham}
\end{equation}
Here $c_{i\sigma}^\dag(c_{i\sigma})$ are the creation (annihilation) operators of an electron at the site $i$ with the spin projection ${\sigma = \pm 1/2}$ on the normal to the 2D plane, $z$ axis, and $\epsilon_i$ are the spin independent site energies. The second term in Eq.~\eqref{eq:ham} describes the spin-dependent hopping with the amplitudes ${J}_{ij}^{\sigma\sigma'}$. $\mathcal H_Z$ is the Zeeman Hamiltonian. In this {and the next} section we neglect the electron $g$-factor for the sake of simplicity thus assuming $\mathcal H_Z=0$. The modification of kinetic coefficients accounting for the Zeeman splitting is discussed in Sec.~\ref{sec:Zeeman}.


The hopping amplitude is determined by the transfer integral
\begin{equation}
  J_{ij}^{\sigma\sigma'}\sim\int\d\bm r  \Psi_\sigma^*(\bm r-\bm r_i) V(\bm r) \Psi_{\sigma'}(\bm r-\bm r_j),
\end{equation}
where $V(\bm r)$ is the potential energy including the attraction potential of sites $i$ and $j$. 
The localized electron wave function has the asymptotic form~\cite{KKavokin-review}
\begin{equation}
{\Psi_\sigma(\bm r)\sim \exp{\left\lbrace\frac{i}{\hbar}\int\limits_0^{\bm r}\left[\bm p(\bm r')-\frac{e}{c}\widetilde{\bm A}(\bm r')\right]\d\bm r'\right\rbrace}\chi_\sigma.
}
\label{eq:Psi}
\end{equation}
Here $\chi_\sigma$ is the basis spinor,
$\bm p(\bm r)$ is the imaginary quasiclassical momentum of electrons, and 
\begin{equation}
{\widetilde{\bm A}(\bm r) = \bm{A}(\bm{r}) - {cm\over\hbar e}\hat{\bm{\sigma}}\cdot \hat{\bm \beta}}
\end{equation}
 is the modified vector potential. It includes the vector potential of the applied magnetic field and the term corresponding to spin-orbit interaction.  
 This allows us to obtain~\cite{Holstein1961,Raikh,Lyanda_PRL,KozubPRL}
\begin{subequations}
  \begin{equation}
  \label{U_ij}
    \hat{J}_{ij}=J_{ij}\hat{U}_{ij},
    \end{equation}
where
    \begin{equation}
        J_{ij}=J_0\e^{-r_{ij}/a_b},
        \quad
\hat{U}_{ij}=\exp\left(-i\bm{d}_{ij} \cdot \bm{\hat{\sigma}}+i\varphi_{ij}\right).
 \end{equation}
Here we neglected power-law terms in $ \hat{J}_{ij}$ in comparison to the exponential dependence $e^{-r_{ij}/a_b}$.
The spin-orbit and magnetic-field induced phases are given by
  \begin{equation}
    \bm{d}_{ij}={m\over \hbar^2} \hat{\bm \beta}\bm{r}_{ij},
    \quad
    \varphi_{ij}=\frac{e\bm B}{2\hbar c}\cdot\left(\bm r_i\times\bm r_j\right),
  \end{equation}
\end{subequations}
where $m$ is the electron effective mass, $\bm r_i$ are the coordinates of the sites in the 2D plane, $\bm{r}_{ij}=\bm{r}_i-\bm{r}_j$, $a_b$ is the localization length~\cite{Efros89_eng} and we have used the Coulomb gauge. $J_0$ is a real constant of the order of the binding energy. In general case $J_0$ and $a_b$ are even functions of the magnetic field~\cite{Efros89_eng}.

The phonon Hamiltonian has the form
\begin{equation}
   {\cal H}_{ph}=\sum_{\bm q}\hbar \Omega_{\bm q}b_{\bm q}^\dag b_{\bm q},
\end{equation}
where $\hbar\Omega_{\bm q}$ is the energy of the phonon with the wavevector $\bm q$, and $b_{\bm q}(b_{\bm q}^\dag)$ is the corresponding annihilation (creation) operator. The Hamiltonian of the electron-phonon interaction reads
\begin{equation}
   {\cal H}_{e-ph}=\sum_{i,\sigma,\bm{q}}v_q(\e^{\i\bm{qR_i}}b_{\bm q}+\e^{-\i\bm{qR_i}}b_{\bm q}^\dag)c_{i\sigma}^\dag c_{i\sigma}
\end{equation}
with $v_q$ being the electron-phonon interaction constants. The spin dependence of electron-phonon interaction is negligible.

After the canonical transformation~\cite{Polarons,BryksinReview}, the total Hamiltonian can be presented as
\begin{equation}
  {\cal H}=\sum_{i,\sigma}\epsilon_ic_{i\sigma}^\dag c_{i\sigma}+\sum_{\bm q}\hbar\Omega_qb_{\bm q}^\dag b_{\bm q}+\sum_{i,j,\sigma,\sigma'}{V}_{ij}^{\sigma\sigma'}c_{i\sigma}^\dag c_{j\sigma'},
\end{equation}
where $V_{ij}^{\sigma\sigma'}={J}_{ij}^{\sigma\sigma'}Q_{ij}$ with
\begin{equation}
  Q_{ij}=\exp\left\lbrace-\sum_{\bm q}\gamma_q\left[\left(\e^{\i\bm{qr}_i}-\e^{\i\bm{qr}_j}\right)b_{\bm q}+\rm{h.c}\right]\right\rbrace,
\label{eq:Vij}
\end{equation}
and $\gamma_q=v_q/(\hbar\Omega_q)$.

The aim of microscopic theory is to
to derive the kinetic equation.
For the sake of simplicity we limit ourselves
to the lowest orders of perturbation theory in the electron-phonon and spin-orbit interactions.
%
Provided the concentration of charge carriers $n$ is inferior by far than the concentration of localization sites $n_s$, one can neglect correlation effects. Additionally, under assumption that the concentration of localization sites is
much smaller
than the localization length $a_b$, $n_sa_b^2\ll 1$, we will use the on-site spin  density matrices $\hat{\rho}_i$.
The master equation can be  presented as
\begin{equation}
  \frac{\d\hat{\rho}_i}{\d t}=\sum_k\frac{\d\hat{\rho}_i^{(k)}}{\d t},
\end{equation}
where the sum runs over the orders of perturbation theory in the hopping amplitude.

The first nonvanishing term is the second-order contribution
\begin{multline}
  \frac{\d{\hat\rho}_i^{(2)}}{\d t}=\sum_{m}\frac{\pi}{\hbar} \left\langle\delta(E_n-E_m)\left(2\hat{V}_{nm}\hat\rho_j\hat{V}_{mn}
\right.\right.\\\left.\left.
-\hat{\rho}_i\hat{V}_{nm}\hat{V}_{mn}-\hat{V}_{nm}\hat{V}_{mn}\hat{\rho}_i\right)\right\rangle.
\end{multline}
Here $n$ and $m$ denote the states of the electron-phonon system where the given electron is localized at sites $i$ and $j$, respectively, and the angular brackets denote averaging over the phonon bath state. The contribution ${\hat\rho}_i^{(2)}$ describes hopping and spin rotations.
Since
\begin{equation}
  \hat{V}_{nm}\hat{V}_{mn}=J_{ij}^2\hat{1} Q_{ij}Q_{ji},
\end{equation}
where $\hat{1}$ denotes the  $2\times2$ unit matrix,
the outgoing term
in the second order
is the same for all spin orientations. Accordingly the second-order contribution can be presented as
\begin{equation}
  \frac{\d{\hat\rho}_i^{(2)}}{\d t}=\sum_j\left(-\frac{{\hat\rho}_i}{\tau_{ji}}+\frac{\hat U_{ij}{\hat\rho}_j\hat{U}_{ji}}{\tau_{ij}}\right),
\label{eq:rho2U}
\end{equation}
with $\hat{U}_{ij}$ being the unitary spin rotation operator Eq.~\eqref{U_ij}
and $\tau_{ji}$ being the hopping time from the site $i$ to  $j$.
%
In the lowest (second) order in the electron-phonon interaction the hopping time is given by
\begin{equation}
  \frac{1}{\tau_{ji}}=
\frac{2\pi}{\hbar}J_{ij}^22\gamma_{q_{ij}}^2 D(\left|\epsilon_{ij}\right|)\left[N_{|\epsilon_{ij}|}+\Theta(\epsilon_{ij})\right],
\label{eq:tau}
\end{equation}
where $\epsilon_{ij}=\epsilon_i-\epsilon_j$, $q_{ij}$ is the phonon wave vector corresponding to this energy, $\Theta(\epsilon)$ is the Heaviside function, $D(\epsilon)$ stands for the {phonon density of states}, and ${N_\epsilon=1/\left[\exp(\epsilon/k_B T)-1\right]}$ is the occupation of the phonon state
with $k_B$ and $T$ being the Boltzmann constant and temperature, respectively. This result can be conveniently obtained using the hopping diagrams introduced in Ref.~\cite{Hopping_spin}. The multiplier $2$ reflects the fact that the phonon can be emitted either at site $i$ or $j$. Note that, due to the energy difference, $\tau_{ji}\neq\tau_{ij}$. As it is commonly accepted, we neglect simultaneous hops of two and more electrons.

In what follows we derive all {the other} terms of the kinetic equation in the lowest nonvanishing order in spin-orbit interaction. The effective frequency of spin precession during the hop is accordingly given by $\bm{\Omega}_{ij}=2\bm{d}_{ij}/\tau_{ij}$ provided $\left|\bm{\Omega}_{ij}\tau_{ij}\right| \ll 1$.
%

{The} third-order contribution to the master equation has the form
\begin{widetext}
\begin{equation}
  \frac{\d{\hat\rho}_i^{(3)}}{\d t} =-\frac{4\pi}{\hbar}\sum_{m,l}\left\langle \delta(E_n-E_m) \left\{
\pi\Im\left(\hat V_{nm}\hat \rho_j\hat V_{ml}\hat V_{ln}\right)\delta(E_n-E_l)
+\frac{\Re\left[\hat \rho_i\Re\left(\hat V_{nm}\hat V_{ml}\hat V_{ln}\right) - \hat V_{nm}\hat \rho_j\hat V_{ml}\hat V_{ln}\right]}{E_n-E_l}\right\} \right\rangle,
\end{equation}
where we have introduced the notations $\Re O\equiv(O+O^\dag)/2$ and $\Im O\equiv(O-O^\dag)/(2i)$.
It {can be rewritten in a form similar to Eq.~\eqref{eq:rho2U}:}
\begin{equation}
  \label{eq:rho3}
  \frac{\d{\hat\rho}_i^{(3)}}{\d t} =\sum_{jk}\left\lbrace\Re\left[\frac{\hat U_{ij}\hat \rho_j\hat U_{jk}\hat U_{ki}}{\tau_{ikj}}-\frac{\hat \rho_i\Re\left(\hat U_{ij}\hat U_{jk}\hat U_{ki}\right)}{\tau_{jki}}\right]-\frac{\Im\left(\hat U_{ij}\hat \rho_j\hat U_{jk}\hat U_{ki}\right)}{\tau_{ikj}'}\right\rbrace,
\end{equation}
\end{widetext}
where
\begin{subequations}
  \label{eq:tau3}
  \begin{equation}
    \frac{1}{\tau_{ikj}}=\frac{1}{\tau_{ij}}\frac{J_{ik}J_{kj}}{2J_{ij}}\left(\frac{1}{\epsilon_i-\epsilon_k}+\frac{1}{\epsilon_j-\epsilon_k}\right),
    \label{eq:tau3odd}
  \end{equation}
  \begin{equation}
    \frac{1}{\tau_{ikj}'}=\frac{\hbar}{4}\left(\frac{J_{ij}}{J_{kj}J_{ki}\tau_{ik}\tau_{kj}}+\frac{J_{jk}}{J_{ij}J_{ki}\tau_{ki}\tau_{ij}}+\frac{J_{ik}}{J_{ij}J_{kj}\tau_{ij}\tau_{kj}} \right).
  \end{equation} 
\end{subequations}
These expressions can be also directly obtained from the diagrammatic approach~\cite{Hopping_spin}. We note that the rate $1/\tau_{ikj}$ describes emission/absorption of one phonon.
{These rates contribute to} the interference mechanism of magnetoresistance~\cite{Shklovskii_Spivak,Shumilin_Kozub}.
The rate $1/\tau_{ikj}'$ describes interaction with at least two phonons. 
The corresponding processes lead to the hopping Hall effect~\cite{Holstein1961,Galperin}.
{For our purposes, it is important to keep both contributions because they} have different symmetry.

It is convenient to present the on-site density matrix in the form
  \begin{equation}
    \hat{\rho}_i= {n_i\over 2} \hat{1}+\hat{\bm{\sigma}}\cdot \bm S_i,
    \label{eq:rho_s}
  \end{equation}
where $n_i$ is the occupancy of site $i$, and $\bm S_i$ is the corresponding spin density.
Substitution {of this expression into} Eq.~\eqref{eq:rho3} yields a system of coupled kinetic equations:
  \begin{subequations}
    \label{eq:kinetic}
  \begin{equation}
    \dot{n}_i=\sum_j I_{ij}+\sum_{j}\left(\bm{\Lambda}_{ij} \cdot \bm{S}_j-\bm{\Lambda}_{ji} \cdot \bm{S}_i\right),
\label{eq:kinetic_n}
  \end{equation}
\begin{multline}
\dot{\bm S}_i +\sum_j \bm{S}_{j}\times \bm{\Omega}_{ij}+\frac{\bm S_i}{\tau_s} + {\bm S}_i \times \bm \Omega_L \\ = \sum_{j} \bm I_{ij}^s +
\sum_{j} \left({\bm G}_{ij} n_j+{\bm G}_{ji} n_i\right).
\label{eq:kinetic_s}
  \end{multline}
\end{subequations}
Here
\begin{equation}
I_{ij}= {n_j \over \tau_{ij} } - {n_i \over \tau_{ji} }
\end{equation}
is the particle flow between sites $i$ and $j$,
and $\bm\Omega_L$ is the Larmor precession frequency in the external magnetic field.
Assuming that spin relaxation is mainly governed by the on-site hyperfine interaction, we phenomenologically introduced the spin relaxation time $\tau_s$.
We note that
the hopping time $\tau_{ij}$ as well as the spin relaxation time $\tau_s$ can be anisotropic, which is disregarded in Eqs.~\eqref{eq:kinetic}.
The spin current flowing from the site $j$ to the site {$i$ is} a sum of two contributions
\begin{equation}
{\bm{I}}^s_{ij}=\frac{\bm{S}_j}{\tau_{ij}}-\frac{\bm{S}_i}{\tau_{ji}}
	+ \bm{W}_{ij} n_j-\bm{W}_{ji} n_i.
\label{eq:Is}
\end{equation}
The first two terms describe spin diffusion, while the latter terms arise due to a difference in spin-conserving tunneling rates for electrons with spin oriented along ($\uparrow$) and opposite ($\downarrow$) to the axis $\alpha$: $W_{ij}^\alpha=(W_{\uparrow\uparrow}-W_{\downarrow\downarrow})/2$. Similarly  $G_{ij}^\alpha=\left(W_{\uparrow\downarrow}-W_{\downarrow\uparrow}\right)/2$ describes spin generation. The spin-galvanic coefficient can be presented as ${\Lambda_{ij}^\alpha=}2\left(W_{\uparrow\uparrow}+W_{\downarrow\uparrow}-W_{\downarrow\downarrow}-W_{\uparrow\downarrow}\right)$. Therefore we obtain a general relation
  \begin{equation}
     	\bm{\Lambda}_{ij} = 4\left(\bm{W}_{ij}-\bm{G}_{ij}\right).
        \label{eq:relation}
  \end{equation}

The kinetic coefficients $\bm{\mathcal K}_{ij}$ ($\bm{\mathcal K} = \bm \Lambda, \bm G, \bm W $) {in Eq.~\eqref{eq:kinetic}} are equal to sums over the auxiliary sites $\bm{\mathcal K}_{ij} = \sum\limits_k \bm{\mathcal K}_{ijk}$, and the relation~\eqref{eq:relation} holds for $\bm{\mathcal K}_{ijk}$ as well. These expressions
demonstrate that CISP, SGE and SHE arise only taking into account hopping between three sites, i.e. triads should be considered.
From the ingoing contributions in Eq.~\eqref{eq:rho3}
we obtain that
\begin{subequations}
  \label{eq:gamma_lambda}
  \begin{multline}
    \label{eq:gamma}
    \bm{\Gamma}_{ikj}\equiv\bm{G}_{ikj}+\bm{W}_{ikj}=\alpha_{xy}\alpha_{yx}\biggl[\frac{2}{3}\bm{A}_{ikj}\times\hat{\bm \alpha}\left(\bm{r}_{ij}+\bm{r}_{ik}\right)
\\
    -\bm{A}_{ikj}\biggr]\left(\frac{\cos{\varphi_{ikj}}}{\tau_{ikj}'}+\frac{\sin{\varphi_{ikj}}}{\tau_{ikj}}\right),
  \end{multline}
  \begin{multline}
    \label{eq:lambda}
    \bm{\Lambda}_{ikj}=4\alpha_{xy}\alpha_{yx}\left[\frac{2}{3}\bm{A}_{ikj}\times\hat{\bm \alpha}\left(\bm{r}_{jk}+\bm{r}_{ji}\right)
    -\bm{A}_{ikj}\right]
\\
\times\left(\frac{\cos{\varphi_{ikj}}}{\tau_{ikj}'}+\frac{\sin{\varphi_{ikj}}}{\tau_{ikj}}\right),
  \end{multline}
\end{subequations}
where $\bm{A}_{ikj}=\bm{r}_{ki}\times\bm{r}_{ij}/2$ is the oriented area of the triad, $\hat{\bm\alpha}=m\hat{\bm\beta}/\hbar^2$ and
\begin{equation}
  \varphi_{ikj}=\varphi_{ij}+\varphi_{jk}+\varphi_{ki}=2\pi\frac{\Phi_{ikj}}{\Phi_0}
\end{equation}
with $\Phi_{ikj}=\bm B \cdot \bm A_{ikj}$ being the magnetic flux through the triad and $\Phi_0=2\pi\hbar c/|e|$ being the magnetic flux quantum,
see inset in Fig.~\ref{fig:magn_field}.
Using the relation~\eqref{eq:relation} one finds
  \begin{subequations}
    \label{eq:G_W}
  \begin{equation}
    \bm{G}_{ikj}=\alpha_{xy}\alpha_{yx}\bm{A}_{ikj}\times\hat{\bm \alpha}\bm{r}_{ij}\left(\frac{\cos{\varphi_{ikj}}}{\tau_{ikj}'}+\frac{\sin{\varphi_{ikj}}}{\tau_{ikj}}\right),
  \end{equation}
  \begin{multline}
 \label{eq:W}
    \bm{W}_{ikj}=\alpha_{xy}\alpha_{yx}\left[\frac{\bm{A}_{ikj}}{3}\times\hat{\bm \alpha}\left(\bm{r}_{jk}+\bm{r}_{ik}\right)   -\bm{A}_{ikj}\right]
\\
\times\left(\frac{\cos{\varphi_{ikj}}}{\tau_{ikj}'}+\frac{\sin{\varphi_{ikj}}}{\tau_{ikj}}\right).
  \end{multline}
  \end{subequations}
We see that the kinetic coefficients oscillate with magnetic field, and the period of oscillations is determined by the triad area $A_{ikj}$.

{We note that the phase related with the spin-orbit interaction is equivalent to the dynamical phase factor:
  \begin{equation}
    \hat{\bm\sigma}\bm d_{ij}=\frac{1}{\hbar}\int\limits\hat{\bm\sigma}\hat{\bm\beta}\bm{k}_{ij}(t)\d t,
  \end{equation}
  where the wavevector $\bm k_{ij}(t)$ describes propagation of an electron from site $j$ to $i$. In the same time, the Aharonov-Bohm phase $\varphi_{ij}$ is known to be geometric or Berry phase~\cite{Topological-review,berry1984quantal}.}

\subsection{General properties of kinetic equation}
\label{sec:kinetic}

Summation of Eq.~\eqref{eq:kinetic_s} over all sites yields the total spin generation rate in the form
\begin{equation}
\label{sum_S_i_dot}
  \sum_i\dot{\bm S}_i={\sum_{ijk}}'\bm\Upsilon_{ikj} + \sum_{ij}\bm\Omega_{ij}\times\bm S_j - \sum_{i}\frac{\bm S_i}{\tau_s},
\end{equation}
where the prime denotes that each pair $(j,k)$ should be taken only once, and
\begin{equation}
  \bm\Upsilon_{ikj}=\bm\Gamma_{ikj}n_j+\bm\Gamma_{ijk}n_k-\frac{n_i}{4}\left(\bm\Lambda_{jki}+\bm\Lambda_{kji} \right).
\end{equation}
Note that the terms with spin conserving tunneling rates ($\bm W_{ij}$) canceling each other after summation are kept in this expression  for convenience.

In thermal equilibrium the rate $\bm\Upsilon_{ikj}$ vanishes, and the spin polarization is absent. This can be explicitly shown with the help of relations:
\begin{equation}
  \label{eq:equilibrium}
  \frac{n_j}{\tau_{ij}}=\frac{n_i}{\tau_{ji}},
  \quad
  \frac{n_j}{\tau_{ikj}}=\frac{n_i}{\tau_{jki}},
  \quad
  \frac{n_j}{\tau_{ikj}'}=\frac{n_k}{\tau_{ijk}'}, 
  \quad
\tau'_{ikj} = \tau'_{kij}.
\end{equation}
The first of these relations follows from Eq.~\eqref{eq:tau} and represents the detailed balance equation $I_{ij}=0$, while the rest follow directly from Eqs.~\eqref{eq:tau3}. These expressions along with the definitions Eq.~\eqref{eq:gamma_lambda} yield
\begin{multline}
  \bm\Lambda_{jki}'+\bm\Lambda_{kji}'=0,
  \quad
  \bm\Gamma_{ikj}'n_j+\bm\Gamma_{ijk}'n_k=0,
  \\
  \bm\Gamma_{ikj}''n_j-\frac{1}{4}\bm\Lambda_{jki}''n_i=0,
  \quad
  \bm\Gamma_{ijk}''n_k-\frac{1}{4}\bm\Lambda_{kji}''n_i=0,
\end{multline} 
where one and two primes denote the even in {$B_z$ contributions}  proportional to $1/\tau_{ikj}'$ and the odd in {$B_z$ ones} proportional to $1/\tau_{ikj}$, respectively, see Eq.~\eqref{eq:gamma_lambda}.
Combining all together one finds $\bm\Upsilon_{ikj}=0$ in thermal equilibrium, as expected.
In close to equilibrium conditions we obtain:
\begin{equation}
  \bm\Upsilon_{ikj}=\bm\Gamma_{ikj}'\tau_{kj}I_{kj}+\bm\Gamma_{ikj}''\tau_{ij}I_{ij}+\bm\Gamma_{ijk}''\tau_{ik}I_{ik}.
  \label{eq:upsilon}
\end{equation}
This expression is similar to ``Hall source'' in the theory of hopping Hall effect~\cite{Galperin}.



The average spin evolution follows from Eq.~\eqref{sum_S_i_dot}:
  \begin{equation}
  \label{dot_s}
    \dot{\bm s} = \frac{1}{nA}\sum_i\dot{\bm S_i} = \frac{1}{nA}\sum_{ij}\left(2\bm G_{ij}n_j+\bm\Omega_{ij}\times\bm S_j\right)
- {\bm s \over \tau_s},
  \end{equation}
where $A$ is the total area of the sample. 
This expression differs from Eq.~\eqref{sum_S_i_dot} by omission of spin conserving tunneling terms.
It can be conveniently rewritten introducing the total spin current
\begin{equation}
  \Js=\frac{1}{2A}\sum_{ij}{\bm r}_{ij}I_{ij}^{s,z},
\label{eq:Js}
\end{equation}
as follows~\cite{Hopping_spin,kkm94}
  \begin{equation}
     \dot{\bm s} = -{2\over n} \bm e_z\times\hat{\bm\alpha}{\bm{\mathcal{J}}} - {\bm s \over \tau_s},
\label{eq:IKKM}
  \end{equation}
with $\bm e_z$ being a unit vector along the $z$ axis.
We remind that we restrict ourselves only to the lowest (third) order in spin-orbit interaction. Defined in this way the spin current vanishes in thermodynamic equilibrium. One can separate two qualitatively different contributions to the spin current: $\Jdiff$ and $\Jdr$, as the two first and two latter terms in Eq.~\eqref{eq:Is}. Provided the electric field is applied to the structure along $x$ direction the difference between two contributions in the perpendicular direction is related only to the spin relaxation:
  \begin{equation}
    \mathcal J_y=\frac{1}{\tau_sA}\sum_i y_{i}S_i^z.
    \label{eq:Jstaus}
  \end{equation}
The spin current in the longitudinal ($x$) direction  can be nonzero even without spin relaxation as a product of spin polarization and electric current.
We remind that, in accordance with the symmetry analysis performed in Sec.~\ref{sec:phenomen}, the odd and even in $B_z$ contributions to spin orientation and spin current averaged over disorder are perpendicular to each other.

It follows from {Eq.~\eqref{dot_s} in the steady state} that the CISP conductivity can be presented as
\begin{equation}
\label{sigma_CISP}
  \hat{\bm \sigma}_\text{CISP}= \left[f(n_s,\tau_s)+g(n_s,\tau_s)\bm e_z\times\right]\Tr(\hat{\bm\beta}^2) \hat{\bm \beta}^T\mathcal{P}\tau_s.
\end{equation}
Here
\begin{equation}
\mathcal{P}=\left({m a_b \over \hbar^2}\right)^3 {2\hbar n_sa_b \over enJ_0\tau_0\rho},
\end{equation} 
$\rho$ is the resistivity, $J_0$ and $\tau_0$ are the characteristic hopping integral and time for the distance $\sim a_b$~\footnote{Here we define $\mathcal{P}$ as a value {two times} larger than in Ref.~\cite{Hopping_spin}.}. The dimensionless functions $f(n_s,\tau_s)$ and $g(n_s,\tau_s)$ are even and odd in $B_z$, respectively, as follows from the symmetry analysis presented in see Sec.~\ref{sec:phenomen}.

The spin-galvanic current can be similarly obtained from the kinetic equation~\eqref{eq:kinetic_n}. The calculation yields the following result for the SGE response:
  \begin{equation}
      \hat{\bm \sigma}_\text{SGE}= \left[f(n_s,\tau_s)-g(n_s,\tau_s)\bm e_z\times\right]4 \Tr(\hat{\bm\beta}^2) \hat{\bm \beta}^T \mathcal{P}k_BT n.
  \end{equation}
Here the functions $f$ and $g$ coincide with those for CISP, Eq.~\eqref{sigma_CISP}, as follows from the Onsager relation~\cite{Levitov1985,Vignale2014,Hopping_spin}.

The spin-Hall conductivity can be deduced from Eqs.~\eqref{eq:IKKM} and~\eqref{sigma_CISP}:
  \begin{equation}
    \hat{\bm\sigma}_\text{SHE}= -\left[f(n_s,\tau_s)+g(n_s,\tau_s)\bm e_z\times\right]\hat{\bm\beta}^T \left(\bm{e}_z\times\hat{\bm\beta} \right)\frac{\hbar^2n\mathcal{P}}{m}.
    \label{eq:sigma_SHE}
  \end{equation}
We stress that, in the inhomogeneous system under study, the drift and diffusion currents are always interconnected. Therefore the spin-Hall conductivity describes the {\it total} spin current induced by the applied electric field. The pure drift spin current, leading to spin separation, can be found formally from Eq.~\eqref{eq:sigma_SHE} in the limit $\tau_s\to0$ when the diffusion spin current vanishes.


\section{Disorder averaging}
\label{sec:averaging}

The above analysis provides microscopic equations that describe CISP, SGE and SHE in the hopping regime. Ultimately, we are interested in the macroscopic susceptibilities introduced in Eqs.~\eqref{eq:3effects}. However in the disordered system the link between microscopic expressions and macroscopic parameters is not straightforward due to an exponential distribution of the hopping times.


Equations~\eqref{sigma_CISP}---\eqref{eq:sigma_SHE} express macroscopic susceptibilities through the dimensionless functions $f(n_s,\tau_s)$ and $g(n_s,\tau_s)$. In this section we study in detail the even in magnetic field effects that are described by $f(n_s,\tau_s)$. In what follows, for brevity we call the function $f(n_s,\tau_s)$ the {\it spin susceptibility}.
As it is shown in the previous section, the kinetic coefficients, Eqs.~\eqref{eq:gamma_lambda} and~\eqref{eq:G_W} oscillate with magnetic field. In this section we demonstrate that these oscillations are strongly modified in a macroscopic system due to the disorder.

%
%

We consider the system with dominant spatial disorder. So we assume that the energy disorder  $|\epsilon_i-\epsilon_j|$ is small or comparable to the temperature. In this case we can neglect the dependence of hopping times $\tau_{ij}$ on energies in comparison to the strong dependence 
on site positions. Finally we limit ourselves to Ohmic regime.
In the analysis of the magnetic-field dependence of the spin susceptibility  we neglect for simplicity the dependencies of $\tau_0$ and $a_b$ on $B_z$ as well as magnetoresistance.

\subsection{Numerical simulation}

\begin{figure}[b]
  \centering
  \includegraphics[width=\linewidth]{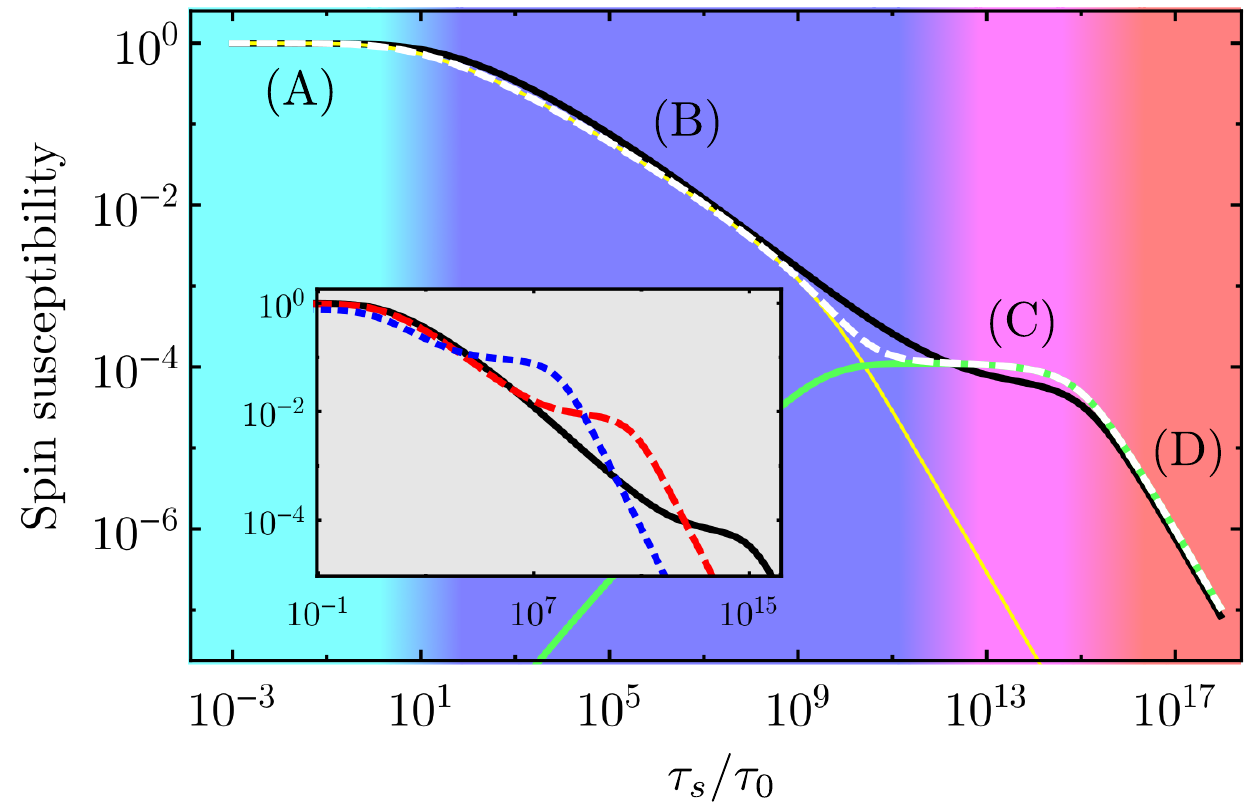}
  \caption{The spin susceptibility $f(n_s,\tau_s)$ calculated numerically for $n_sa_b^2=0.01$ (black curve) and its analytical approximation,  Eq.~\eqref{eq:Jnet}, (white dashed curve) with the parameters $\gamma=1$ and $D=0.5\times 10^{-8}a_b^2/\tau_0$. The background colors distinguish the four regimes (A), (B), (C) and (D) discussed in the text. Yellow and green curves show the two contributions, Eqs.~\eqref{eq:Jdistrib} and~\eqref{eq:J_med_res}, to the spin current in the percolation model. The inset shows the same black solid curve and the spin susceptibility calculated for $n_sa_b^2=0.3$ (red dashed curve) and $0.1$ (blue dotted curve).}
  \label{fig:f}
\end{figure}

\begin{figure*}
  \centering
  \includegraphics[width=\linewidth]{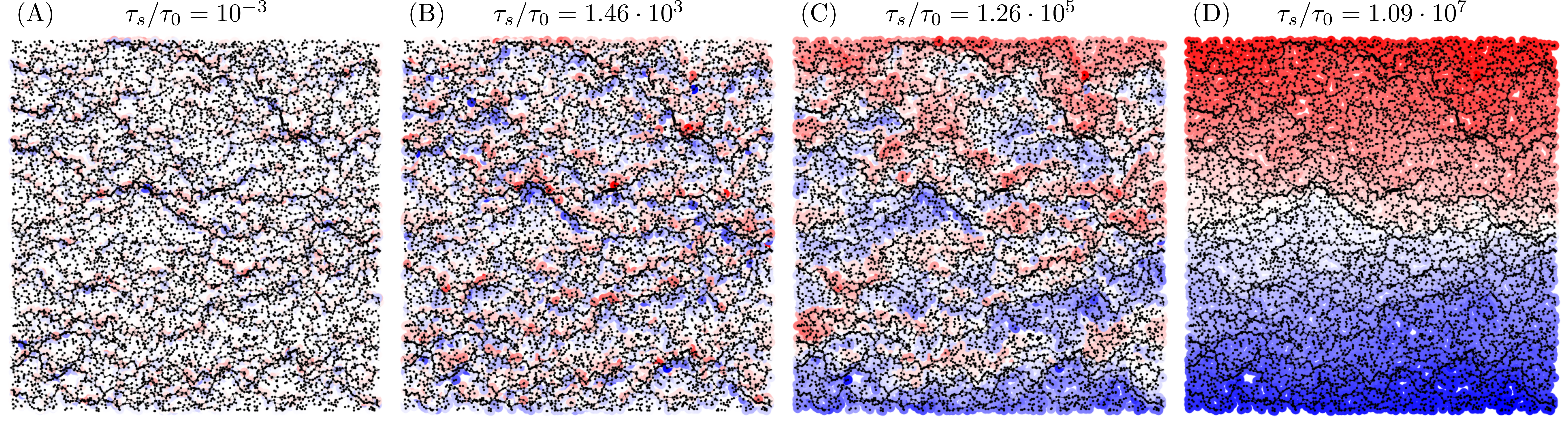}
  \caption{Distributions of $S_{i,z}$ for different spin relaxation times in zero magnetic field. The four panels from left to right correspond to the regimes (A)---(D). The red color correspond to $S_{i,z}>0$ and blue color --- to $S_{z,i}<0$. The color scale is arbitrary. The black lines show the particle fluxes between the sites. Parameters of the calculation are $n_sa_b^2=0.1$ and $N_s=10^4$.}
  \label{fig:maps}
\end{figure*}


We have performed a numerical simulation of coupled charge and spin dynamics described by Eqs.~\eqref{eq:kinetic}. As it is mentioned above, we have assumed that the spatial disorder dominates over energy disorder. In this case the hopping time in each pair $(ij)$ has the form
\begin{equation}
\tau_{ij}=\tau_{ji}=\tau_0\e^{2r_{ij}/a_b}
\label{eq:tau_exp}
\end{equation}
with $\tau_0$ being a constant. The conductivity of the system was analyzed using Miller-Abrahams random resistor network where each pair is replaced by a resistor with the resistivity $R_{ij}=nk_BT\tau_{ij}/(e^2 n_s)$~\cite{Efros89_eng}. In this model, a numerical solution of a set of Kirchhoff equations yields the particle flows $I_{ij}$ for each pair of sites. In the next step the spin generation rates $\bm\Upsilon_{ijk}$ were calculated using Eqs.~\eqref{eq:upsilon}. 
Then the steady-state spin density was found from {Eqs.~\eqref{eq:kinetic_s}.}
At this step we neglect spin generation rate $\bm G_{ij}$ and spin precession $\bm\Omega_{ij}$ because they are proportional to the third power of spin-orbit constants. And finally the spin current was calculated using Eq.~\eqref{eq:Jstaus}. Comparison of the result with Eq.~\eqref{eq:sigma_SHE} yields the spin susceptibility $f(n_s,\tau_s)$.
We have performed numerical simulations for ${N_s=512\times10^3}$ localization sites with the Poisson distribution, and we have checked that the difference between the three realizations of the disorder in less than $1\%$.

The dependence of the spin susceptibility on the spin relaxation time at zero magnetic field is presented in Fig.~\ref{fig:f} for fixed values of the concentration.
The black line in Fig.~\ref{fig:f} shows the function $f(n_s,\tau_s)$ calculated for $n_sa_b^2=0.01$. 
One can distinguish four regimes in the dependence of the spin susceptibility on the spin relaxation time which are shown by different {background} colors in Fig.~\ref{fig:f}. For small $\tau_s$ we find that $f$ tends to $1$ (cyan region, regime A). When $\tau_s$ increases and reaches the blue region, the spin susceptibility decays approximately as $1/\sqrt{\tau_s}$ (regime B). This decrease stops at a certain value, and in the magenta region of $\tau_s$ the spin susceptibility hardly changes (regime C). Finally for large enough spin relaxation time, $f$ decays as $1/\tau_s$ (red region, regime D). As it is shown in the inset, the second (blue) region narrows down with increase of the concentration $n_s$.

We show the numerically calculated distribution of  generated spin in Fig.~\ref{fig:maps}.  It can be seen that in the regime (A) all the generated spin is localized at close pairs with small separations. In the regime (B) the spin is still localized on rare sites but the separation of the up and down spins is larger. In the regime (C) the generated spin covers entire regions of the sample indicating spin diffusion with a finite length $l_s$. Finally in the regime (D) the spin polarization is distributed over the {whole} sample due to the large spin diffusion length $l_s > L$, {with $L=\sqrt{N_s/n_s}$ being the sample size}.

\begin{figure}[h]
  \centering
  \includegraphics[width=\linewidth]{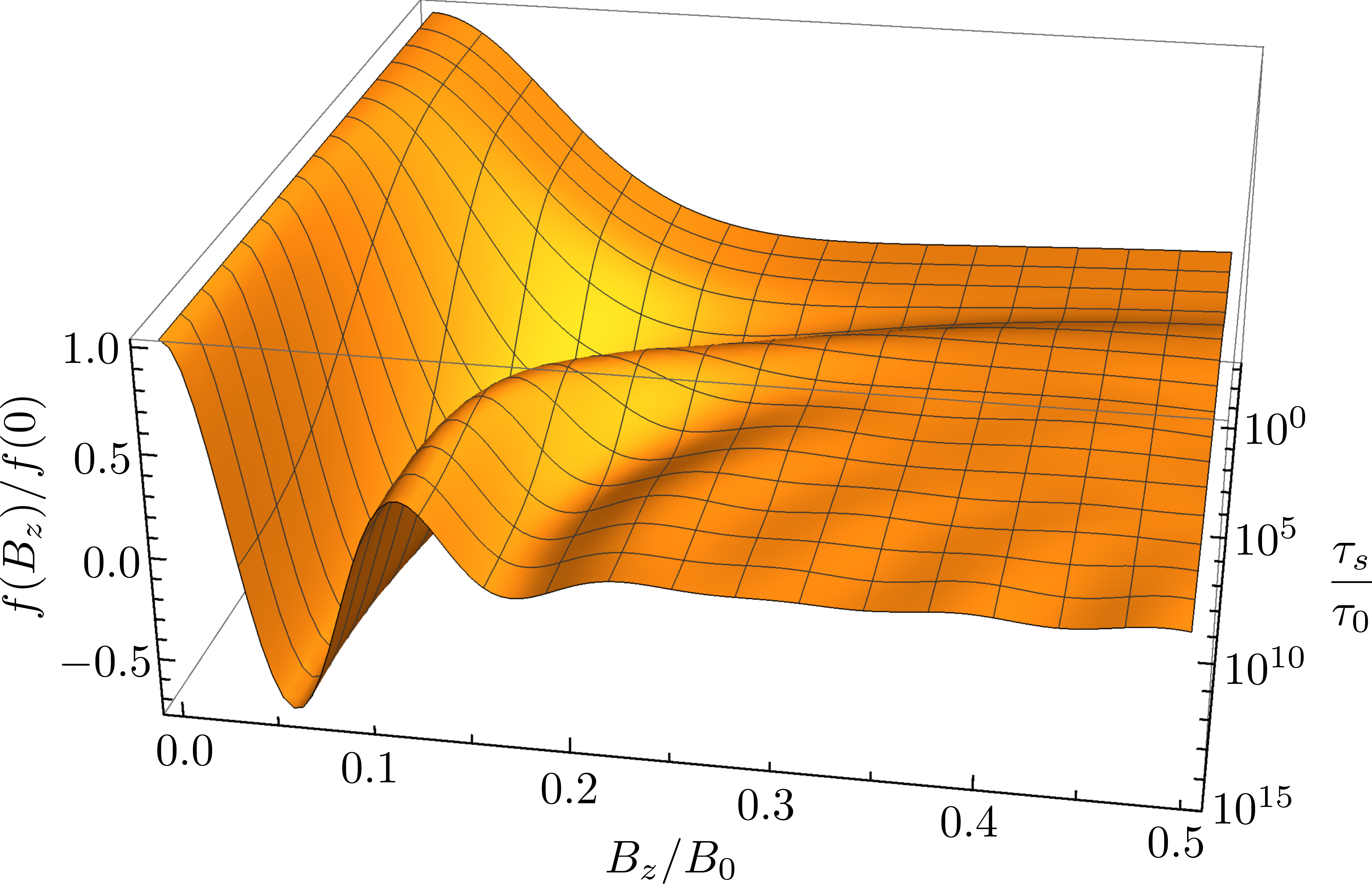}
  \caption{Dependence of the normalized spin susceptibility on $\tau_s$ and $B_z$ for $n_sa_b^2=0.01$.
}
  \label{fig:3D}
\end{figure}

Figure~\ref{fig:3D} demonstrates the magnetic field dependence of the {normalized} spin susceptibility {as a function of two parameters, $\tau_s/\tau_0$ and $B_z/B_0$, where $B_0=\Phi_0/(2\pi a_b^2)$.} One can see that the dependence  $f(B_z)$ can have either only one or multiple changes of sign depending on the relation between the spin relaxation and the hopping times.

\subsection{Percolation analysis}
\label{sec:perc}

In this subsection we develop an analytical theory to describe the dependence of the spin susceptibility on the spin relaxation time and magnetic field. This is possible in the limit of strong disorder, $n_sa_b^2 \ll 1$, when the percolation theory can be applied~\cite{Efros89_eng}.
The presented results are qualitative,
however they are 
in a good agreement 
with numerical simulations.

First, let us briefly summarize the main facts of percolation theory for system conductivity~\cite{Efros89_eng}. As mentioned in the previous subsection, the ensemble of localization sites can be mapped onto the Miller-Abrahams network of resistors with the resistivities
  \begin{equation}
    \label{eq:R_tau}
    R_{ij}\propto\tau_{ij}.
  \end{equation}
Due to the exponentially broad distribution of the hopping times, the current mainly flows in a percolation cluster. It includes only resistors with
  \begin{equation}
    R_{ij} \lesssim R_{perc}= \frac{k_B T n \tau_0}{n_s e^2}\exp(2r_{c}/a_b),
  \end{equation}
 where $r_{c} =2\sqrt{\eta_c/\pi}n_s^{-1/2}$ is the percolation distance. For the 2D system under study the percolation threshold {is} $\eta_c\approx1.128$~\cite{Critical_exponent}.
We note that the system resistivity can be estimated as $\rho\sim R_{perc}$.
Despite the strong disorder, the system can be considered as a homogeneous one with a usual diffusive conductivity on the lengthscale exceeding the correlation length
  \begin{equation}
    L_{cor} = n_s^{-1/2}(r_{c}/a_b)^\nu,
  \end{equation}
with the critical exponent $\nu\approx 1.3$~\cite{Critical_exponent}.


Now we turn to the analysis of the spin susceptibility. 
Similarly to the numerical simulation discussed above, 
its calculation
consists of two steps. In the first step, a distribution of electric currents in the system is determined. In the second step one can analyze the spin-related phenomena on the basis of Eqs.~\eqref{eq:kinetic_s} and~\eqref{eq:upsilon} with the known particle fluxes $I_{ij}$.
The analysis of the spin susceptibility can be conveniently done considering SHE, so we again reduce the kinetic equations~\eqref{eq:kinetic} to the second order in spin-orbit interaction. The corresponding equations for spin dynamics have the form
  \begin{equation}
    \dot{\bf S}_i + \frac{{\bf S}_i}{\tau_s} + \sum_j \frac{{\bf S}_i - {\bf S}_j}{\tau_{ij}} =
    \alpha_{xy}\alpha_{yx} {\sum_{jk}}'{\bf A}_{ikj} \frac{\tau_{jk}\cos\varphi_{ikj}}{\tau'_{ikj}}I_{jk}.
    \label{kin-reduced}
  \end{equation}
Note  that only $z$ component of these equations is nonzero, which corresponds to SHE effect under study.
Here we have neglected the odd in $B_z$ terms in the right hand side because we are aimed only at the description of the even in magnetic field spin susceptibility $f(n_s,\tau_s)$.


The inhomogeneous part of Eqs.~\eqref{kin-reduced} is related to the triads of sites along the percolation cluster where the particle flux is nonzero. Since Eqs.~\eqref{kin-reduced} are linear, the triads can be considered separately. Let us discuss one of these triads $(ijk)$. 
{We separate the contributions to the total spin current, Eq.~\eqref{eq:Jstaus}, from this particular triad, and from all the others, which we model by a diffusive medium as $\tilde S^{(ijk)}(\bm r)$.} It is assumed that the spin polarization can escape each triad with the rate $1/\tau_d$, and  the income of spin polarization from the diffusive medium to the triad under consideration is {negligible}. The corresponding steady-state spin polarizations of the sites satisfy the equations
\begin{align}
    \label{kin_ikj}
      \frac{\tilde S_i^{(ijk)}}{\tau_s'} - \frac{\tilde S_k^{(ijk)}-\tilde S_i^{(ijk)}}{\tau_{ik}} -  \frac{\tilde S_j^{(ijk)}-\tilde S_i^{(ijk)}}{\tau_{ij}} =  I_{kj}\tau_{kj}\Gamma^{(0)}_{ijk}, \nonumber \\
      \frac{\tilde S_j^{(ijk)}}{\tau_s'} - \frac{\tilde S_k^{(ijk)}-\tilde S_j^{(ijk)}}{\tau_{jk}} - \frac{\tilde S_i^{(ijk)}-\tilde S_j^{(ijk)}}{\tau_{ij}} = I_{ik}\tau_{ik}\Gamma^{(0)}_{ijk}, \nonumber \\
  \frac{\tilde S_k^{(ijk)}}{\tau_s'} - \frac{\tilde S_i^{(ijk)}-\tilde S_k^{(ijk)}}{\tau_{ik}} -  \frac{\tilde S_j^{(ijk)}-\tilde S_k^{(ijk)}}{\tau_{jk}} = - I_{ij}\tau_{ij}\Gamma^{(0)}_{ijk},
\end{align}
where $\Gamma^{(0)}_{ijk}=\alpha_{xy}\alpha_{yx}A_{ijk}^z \cos\varphi_{ikj}/\tau'_{ijk}$, 
$${1\over\tau_s'}={1\over\tau_s}+{1\over\tau_d},$$ 
and for the diffusive medium
\begin{multline}
\label{diffusion_eq}
  \frac{\tilde S^{(ijk)}(\bm r)}{\tau_s} - D\Delta\tilde S^{(ijk)}(\bm r)\\=\frac{1}{\tau_d}\left[\tilde S_i^{(ijk)}\delta(\bm r_i)+\tilde S_j^{(ijk)}\delta(\bm r_j)+\tilde S_k^{(ijk)}\delta(\bm r_k)\right]
\end{multline}
with $D$ being the spin diffusion coefficient.
We note that
\begin{equation}
  S_i^{(ijk)}+S_j^{(ijk)}+S_k^{(ijk)}=0,
  \label{eq:zero}
\end{equation}
since we limit ourselves to the study of spin separation and neglect CISP here. The total contribution of the given triad to the total spin current has the form
\begin{equation}
  \mathcal J_{ijk}=\mathcal J_{ijk}^{triad}+\mathcal J_{ijk}^{med},
  \label{eq:J_2contrib}
\end{equation}
where
\begin{equation}
\label{J_spin_triad}
  \mathcal J_{ijk}^{triad}=\frac{1}{\tau_s}\left(y_{i}\tilde S_i^{(ijk)}+y_{j}\tilde S_j^{(ijk)}+y_{k}\tilde S_k^{(ijk)}\right),
\end{equation}
and
\begin{equation}
  \mathcal J_{ijk}^{med}=\frac{1}{\tau_s}\int\d\bm r \: y \: \tilde S^{(ijk)}(\bm r) .
  \label{eq:Jmed}
\end{equation}
The net spin current is presented as
\begin{equation}
  \label{eq:Jnet}
    \mathcal J_y=\frac{1}{A}{\sum_{ijk}}'\mathcal J_{ijk}.
\end{equation}
These equations allow one to describe the dependence of the spin susceptibility on the spin relaxation time and magnetic field shown in Figs.~\ref{fig:f} and~\ref{fig:3D}.

\subsubsection{Zero magnetic field}

First, we analyze the spin susceptibility at zero magnetic field.
The particle flux in each branch of the percolation cluster has the same order of magnitude $I_{perc}$. In a 2D system it can be estimated as $I_{perc} \sim j L_{cor}/e$. In a given triad $I_{perc}$ is divided between the current in pairs $I_{ij}$, $I_{jk}$ and $I_{ki}$ in accordance with the resistivities, Eq.~\eqref{eq:R_tau}.  This defines the right-hand side in Eqs.~\eqref{kin_ikj}. The solution of these equations yields the contribution \eqref{J_spin_triad} of the triad to the total spin current. It turns out that it  has a very strong dependence on the geometry of the triangle formed by the {three} sites under study. The maximum value of this contribution dominates the spin-Hall effect. 

At very short spin relaxation times $\tau_s < \tau_0$, in the regime (A), the maximum is reached in the smallest {triangles, and $\bm {\mathcal J}_{\rm diff}$ can be neglected. Therefore $\bm{\mathcal J}=\bm{\mathcal J}_{\rm dr}$, and the spin susceptibility is} independent of $\tau_s$ in this case. 

For longer spin relaxation times $\tau_s >\tau_0$, regimes (B)---(D), the maximum is reached when the sites $i$, $j$ and $k$ form an equilateral triangle, see Appendix~\ref{App-regimes}. The contribution of the triangle with $r_{ij} = r_{ik} = r_{jk} = r$ to the spin current can be expressed as follows
\begin{equation}
\mathcal J_{0}(r)=- I_{perc}\frac{3\hbar \alpha_{xy}\alpha_{yx}r^3\tau_s'}{16J_0\tau_0\tau_s} \frac{\e^{r/a_b}}{\e^{2r/a_b}+3\tau_s'/\tau_0}.
  \label{eq:J_opt}
\end{equation}
 The side of triangle $r$ is arbitrary in Eq.~\eqref{eq:J_opt}. 
The triad contribution to the {total spin current} can be written in the form
  \begin{equation}
\mathcal J^{triad}=\int_0^{r_c} \d r \: p(r)\mathcal J_{0}(r),
    \label{eq:Jdistrib}
  \end{equation}
where $p(r)$ is proportional to the distribution function of the triangles of the size $r$ along the percolation cluster. We assume that it has the form
\begin{equation}
  p(r)\propto 1/r^\gamma,
  \label{eq:pr}
\end{equation}
where $\gamma$ is a constant.  
This dependence with $\gamma>0$ reflects the fact that the probability to find an equilateral triangle with a side $r \lesssim r_c$ belonging to the percolation cluster drops with $r$.
 
For moderately long spin relaxation times $\tau_s >\tau_0$ {[regime (B)]}, 
the maximum is reached at the optimal value $r=r_{opt}$. 
\begin{equation}
r_{opt}(\tau_s)=\frac{a_b}{2}\ln\frac{3\tau_s}{\tau_0},
\label{eq:ropt}
\end{equation}
where we neglect the contribution $\propto r^3$ in Eq.~\eqref{eq:J_opt} in comparison with {the} fast exponents. The optimal side $r_{opt}$ {is a} result of the interplay of two factors. On one hand, for very large triangles the spin generation efficiency $\Gamma^{(0)}_{ijk}$ decreases exponentially. On the other hand, for small triangles the diffusion  and the drift spin currents exponentially well compensate each other~\cite{Hopping_spin}. In other words, the spin polarization in different directions {at} different sites ``recombines'' due to fast hopping. As a result, there is an exponentially sharp maximum for optimal triangles: $\mathcal J^{triad} \approx \mathcal J_0(r_{opt})$,
and the exact value of $\gamma$ is not very important in comparison with the strong exponential dependence $\mathcal J_0(r)$.

The time $\tau_d$ corresponding to  start of diffusion is related to hopping on the critical distance $r_c$, 
\[ \tau_d \approx \tau_0 \exp(2r_c/a_b).
\] 
The larger is the spin relaxation time $\tau_s$, the larger is the optimal triangle $r_{opt}$. Provided $r_{opt}<r_c$, the diffusive medium in our model does not play an essential role because the generated spin relaxes faster than {$\tau_d$. Therefore the contribution $\mathcal J^{med}$} can be neglected, and the total spin current  $\mathcal J \approx \mathcal J^{triad}$. 
As a result we obtain for regime (B)
\begin{equation}
  \mathcal J_y\propto \mathcal J_0\left[r_{opt}(\tau_s)\right]\propto 1/\sqrt{\tau_s}.
  \label{eq:sqrt}
\end{equation}


In the regimes (C) and (D) the size of the optimal triangle $r_{opt}(\tau_s)$ is larger than the critical distance $r_c$. In this case the main contribution to $\mathcal J^{triad}$ is given by the largest triad along the percolation cluster. At the same time, the spin polarization is partially transferred to the diffusive medium. It follows from Eq.~\eqref{diffusion_eq} that the contribution to the spin current from the diffusive medium has the form
  \begin{equation}
    \mathcal J_{ijk}^{med}=\tilde S_i^{(ijk)}F(y_{i})+\tilde S_j^{(ijk)}F(y_{j})+\tilde S_k^{(ijk)}F(y_{k}),
  \end{equation}
where
\begin{equation}
  F(y)=\frac{1}{\tau_s\tau_d}\int_{-L/2}^{L/2}K(y',y)y'\d y'
  \label{eq:F}
\end{equation}
with
\begin{equation}
  K(y',y)=\frac{\tau_s}{l_s}\frac{\cosh\left(\dfrac{L-|y-y'|}{l_s}\right)+\cosh\left(\dfrac{y+y'}{l_s}\right)}{2\sinh\left(L/l_s\right)}
\end{equation}
being the Green function of the diffusion equation. Here {$|y|<L/2$} with $L$ being the sample length, and ${l_s=\sqrt{D\tau_s} }$ is the spin diffusion length.
Substitution of this expression into Eq.~\eqref{eq:F} yields
\begin{equation}
\label{eq_F}
  F(y)=\frac{1}{\tau_d}\left[y-l_s\frac{\sinh(y/l_s)}{\cosh(L/2l_s)} \right].
\end{equation}
The sizes of triangles $(ikj)$ are smaller than $l_s$ in regimes (C) and (D). This allows us to relate the contribution $\mathcal J_{ijk}^{med}$  to $\mathcal J_{ijk}^{triad}$: $\mathcal J_{ijk}^{med} = \tau_s \mathcal J_{ijk}^{triad} dF/dy$, where we have taken into account Eq.~\eqref{eq:zero}. The contribution $\mathcal J^{med}$ from all the triads is
  \begin{equation}
   \mathcal J^{med}= \mathcal J^{triad}\frac{\tau_s}{\tau_d}\left[1-\frac{2 l_s}{L}\tanh\left(\frac{L}{2l_s}\right)\right].
    \label{eq:J_med_res}
  \end{equation}
Here the multiplier $\tau_s/\tau_d$ describes the ratio of the times spent by the spin inside the triad and outside of it.
In the regime (C) one has $\tau_s\ll L^2/D$ {($l_s\ll L$)}, so the mesoscopic effects do not take place. 
In this case the second terms in Eqs.~\eqref{eq_F} and~\eqref{eq:J_med_res} can be neglected,
and $\mathcal J_y \approx \mathcal J^{med}$ {is} independent of $\tau_s$.
However in the regime (D) the spin separation in the sample {is} suppressed due to diffusion of spin polarization from one boundary of the sample 
to the opposite one, Fig.~\ref{fig:maps}. 
In this regime for $l_s \gg L$ we obtain 
\begin{equation}
  {\mathcal J_y \approx \mathcal J^{med} \propto 1/\tau_s.}
  \label{eq:inverse}
\end{equation}

\subsubsection{Nonzero magnetic field}

Now we proceed to the analysis of the spin susceptibility as a function of an external magnetic field. This dependence is related to the factor $\cos\varphi_{ikj}$ in 
Eq.~\eqref{kin-reduced} which means that the spin separation and spin generation rates in each triad of sites oscillate as functions of $B_z$. Hence one can expect the oscillations of the spin susceptibility similar to Aharonov-Bohm oscillations. Numerical calculation indeed demonstrates this effect, as shown in Fig.~\ref{fig:3D}. 
The presence of oscillations is determined by the spread of oscillations period in optimal triads. If the spread of triad areas is much smaller than the mean area, then the period of Aharonov-Bohm oscillations in a macroscopic system is well defined. Otherwise the oscillations are efficiently smeared.

In the regime (A) the optimal triads are the isosceles triangles with one small side $r_a \sim a_b$, see Appendix~\ref{App-regimes}. The long sides of the triangle $r_{side}$ can be arbitrary large. However we assume that these long sides participate in the percolation cluster, {$r_{side}<r_c$}.  The contribution of such a triangle to the spin current $\mathcal J_{ijk}$
in the regime (A) is
\begin{multline}\label{JS-B-iso}
\mathcal J_{iso} = \frac{3\hbar \alpha_{xy} \alpha_{yx} I_{perc}}{16 J_0 \tau_0} r_{side}  r_{a}^2\cos{\theta} e^{-r_{a}/a_b} \\
 \times \cos \left( 2\pi \frac{B_z r_{side} r_{a}}{2\Phi_0} \right).
\end{multline}
Here $\theta$ is an angle between the long sides of the triangle and the $x$ axis and we have taken into account that $r_a\ll r_{side}$.
This contribution exponentially drops with $r_a$, therefore the area of the optimal triangle can be arbitrary small.  
According to Eq.~\eqref{eq:Jnet}, this expression should be averaged over different optimal triangles to obtain the magnetic field dependence of spin susceptibility. 
Averaging over $\theta$ yields a factor on order of unity.
The distribution of the short sides $r_a$ is related to the probability to find a third site $k$  participating in the percolation cluster near one of the sites $i$ or $j$. The third site $k$ should form an {approximately} isosceles triangle with sites $i$ and $j$, $|r_{ij} - r_{jk}| \lesssim a_b$. The probability to find this site can be estimated as $n_s a_b d r_a$. Integration of Eq.~\eqref{JS-B-iso} with this probability yields
the contribution to the spin current of the isosceles triangles averaged over $r_a$
in the form
\begin{multline}\label{Ap2-2}
\left\langle \mathcal J_{iso} \right\rangle_{r_a}  \sim \\ \frac{\hbar \alpha_{xy}\alpha_{yx} I_{perc} n_s a_b^4 r_{side}}{J_0 \tau_0}
\frac{1-3 \left( {\pi B_z a_b r_{side}/\Phi_0} \right)^2  }{{\left[1 + \left(\pi B_z a_b r_{side}/\Phi_0 \right)^2\right]^3}}.
\end{multline}
The total spin current in the regime (A) is given by averaging of this expression over $r_{side}$. 
The distribution of distances $r_{side}$ between sites in the percolation cluster is not uniform. {When} $r_{side} \ll n_s^{-1/2}$ it can be estimated as $p_A(r_{side}) = p_0 r_{side}$ where $p_0$ is a constant.
This distribution reflects the fact that the probability to find a small triangle with $r_{side}\ll r_c$ in the percolation cluster raises with $r_{side}$.
We extrapolate this distribution up to the largest possible $r_{side} = r_c$. It leads to  
the following expression 
for the spin current in the regime (A) 
\begin{equation}\label{f-B-regA}
f(B_z) =f(0) \: \frac{3}{x^3} \left[ \frac{x+2x^3}{(1+x^2)^2} -  \arctan{x} \right],
\end{equation}
where {$x=B_z r_c/(2B_0 a_b)$.}
The function~\eqref{f-B-regA} does not oscillate but it contains one change of sign. 
%

In the regimes (B)---(D), as discussed above, the optimal triads form equilateral triangles, {see also} Appendix~\ref{App-regimes}. An exponentially sharp maximum exists in the dependence $\mathcal J_0(r)$ meaning that the dominant contribution to SHE comes from the triangles with the same area. With account for the Aharonov-Bohm phase
\begin{equation}
{\cos{\left( {\pi B_z r^2\sqrt{3}\over 2\Phi_0} \right)},}
\end{equation}
we evaluate the integral Eq.~\eqref{eq:Jdistrib} by the stationary-phase method and obtain 
the magnetic-field dependence of the spin susceptibility in the form
\begin{equation}
  f(B_z) = f(0)\cos\left(\frac{B_z}{B_{opt}}\right)
\exp \left( -\frac{2B_z^2}{B_{opt} B_0}\right).
 \label{eq:many_osc}
\end{equation}
Here the period of the oscillations is determined by the area of the optimal triangle:
\begin{equation}
\label{B_opt}
B_{opt} = {4\hbar c\over\sqrt{3}|e|r_{opt}^2},
\end{equation}
and the rate of oscillations decay is related to the decrease of the triad contribution to the spin current when its {size} deviates from the optimal one. {Qualitatively the number of oscillations is of the order of $\sqrt{B_0/B_{opt}}$.}

\section{Discussion}
\label{sec:Zeeman}

The results of the previous section indicate that the dependence of the spin susceptibility on $\tau_s$ as well as its oscillations as a function of the  magnetic field are closely related to the spin transport in strongly disordered sample.

The sum of two contributions, Eqs.~\eqref{eq:Jdistrib} and~\eqref{eq:J_med_res}, describe the total spin current 
in the framework of the percolation analysis at zero magnetic field for any $\tau_s$. The corresponding calculation of the spin susceptibility $f(\tau_s,n_s)$ is shown by the white line in Fig.~\ref{fig:f}. {Reasonably} good agreement of the percolation analysis with the results of numerical calculations is {evident for all the} regimes. Moreover, the analytical dependencies $1/\sqrt{\tau_s}$ for regime (B) {[Eq.~\eqref{eq:sqrt}]} and $1/{\tau_s}$ for regime (D) {[Eq.~\eqref{eq:inverse}]} as well as constants for regimes (A) and (C)
describe the numerical simulations with high accuracy. The contributions to the spin current from  triads and from the {diffusive} media are shown in Fig.~\ref{fig:f} by yellow and green lines, respectively. Figure~\ref{fig:f} demonstrates that the triads' contribution dominates in the regimes (A) and (B). In contrast, triads serve only as sources of the spin current in the regimes (C) and (D) where the diffusive media contribution is {the largest}.

{We note, however that the diffusion coefficient $D=0.5\cdot10^{-8}a_b^2/\tau_0$ used in the analytical calculation in Fig.~\ref{fig:f} is different from the charge diffusion coefficient, obtained from the numerical simulation of system conductivity $5.6\cdot10^{-5}a_b^2/\tau_0$, and from the estimation $L_{cor}^2/\tau_d\approx 2.9\cdot10^{-6}a_b^2/\tau_0$. This is most probably an artifact of our oversimplified model.}

\begin{figure}[h]
  \centering
  \includegraphics[width=\linewidth]{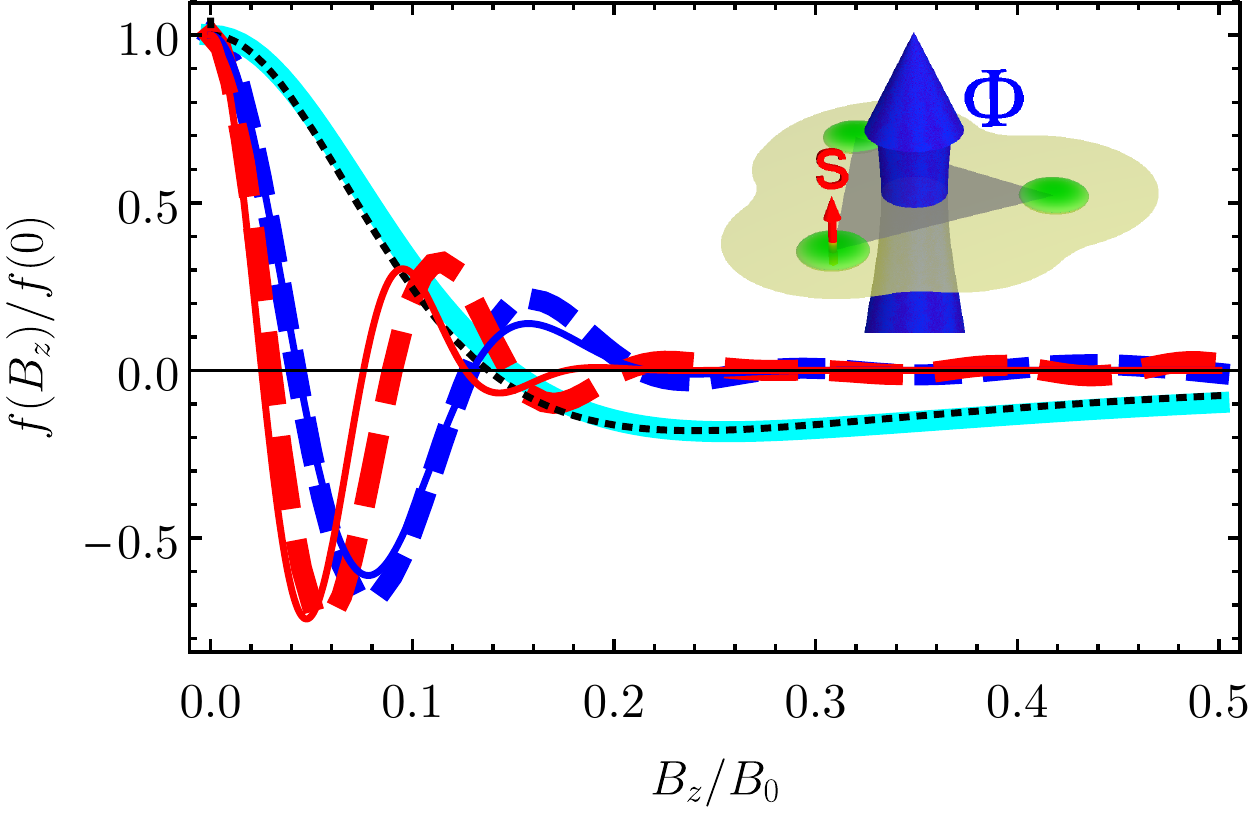}
  \caption{Magnetic field dependencies of the normalized spin susceptibility for $\tau_s/\tau_0=10^{-3}$ (cyan curve), $3\times10^7$ (blue dashed curve) and $4\times 10^{14}$ (red dashed curve). {Parameters of the calculation are the same as in Fig.~\ref{fig:f}.} The numerical results for the susceptibility are compared with Eq.~\eqref{f-B-regA} (dashed black curve) and with Eq.~\eqref{eq:many_osc} (solid blue and red curves). The inset illustrates the magnetic flux through a triad of localization sites responsible for Aharonov-Bohm like oscillations.}
  \label{fig:magn_field}
\end{figure}

Figure~\ref{fig:magn_field} demonstrates the magnetic field dependence of the spin susceptibility for the regimes (A)---(D).
The colors of the curves correspond to the background colors in Fig.~\ref{fig:f}. We note that the magenta curve in the figure is absent, because it coincides with the red one.
The dependence Eq.~\eqref{f-B-regA} is shown by the black dashed curve in Fig.~\ref{fig:magn_field}.
The very good agreement between Eq.~\eqref{f-B-regA} and numerical simulation results in the regime~(A) is clearly seen.
The numerical results for the regimes (B)---(D) agree qualitatively with Eq.~\eqref{eq:many_osc} as shown by solid blue and red curves in Fig.~\ref{fig:magn_field}.  Moreover, the analytical expression~\eqref{B_opt} for the oscillation period is in quantitative agreement with numerical results in the regime (B). In the regimes (C) and (D) the agreement is slightly less perfect: for $\tau_s/\tau_0 =4\times 10^{14}$ 
the numerical result for the period exceeds the analytical estimate Eq.~\eqref{B_opt} by $\sim 18$~\%.

Because of suppression of spin polarization with increase of magnetic field, we focused mainly on the even in magnetic field effects which are described by the spin susceptibility $f(n_s,\tau_s)$. The odd in $B_z$ kinetic coefficients contain energy differences between initial and intermediate states,  Eq.~\eqref{eq:tau3odd}. Therefore they can not be analyzed neglecting energy disorder, as it is done in Sec.~\ref{sec:averaging}. We note, however, that these terms can vanish due to this averaging, which deserves a separate study.

We note that the definitions of macroscopic susceptibilities, Eqs.~\eqref{sigma_CISP}---\eqref{eq:sigma_SHE}, 
are valid for the particular form of spin-orbit interaction, Eq.~\eqref{H_SO}, which is realized in zinc-blende heterostructures grown along $[001]$ direction. Nevertheless the presented results can be applied to a wider class of systems, where one can choose the reference frame in the spin space formally coinciding with Eq.~\eqref{H_SO}. This can be done, in particular, for asymmetric structures grown along [110] direction.
Moreover, despite all three effects in (001) heterostructures are related with the in-plane spin components,  in (110) quantum wells 
the  electric current orients the spin component normal to the 2D plane.
If the (110) system is structure-asymmetric, then its point symmetry group is C$_s$ with a reflection in the  ($yz$) plane being only one nontrivial symmetry element.  Here $z\parallel [110]$ is the normal direction, and $x\parallel[\bar{1}10]$, $y\parallel[001]$ are the in-plane axes~\cite{golub_ganichev_BIASIA}. The symmetry analysis shows that the following even in $B_z$ components are nonzero:
\begin{equation}
\sigma_\text{CISP,SGE,SHE}^{xy,yx}, \qquad
  \sigma_\text{CISP}^{zx},
  \qquad
  \sigma_\text{SGE}^{xz},
\end{equation}
as well as the following odd in $B_z$ ones
\begin{equation}
\sigma_\text{CISP,SGE,SHE}^{xx,yy}, \qquad
  \sigma_\text{CISP}^{zy},
  \qquad
  \sigma_\text{SGE}^{yz}.
\end{equation}
Due to low symmetry, all these components are linearly independent.


In this paper we neglected Zeeman effect, which does not affect the spin current. However, external magnetic field can  significantly suppress the in-plane spin polarization due to Hanle effect {as $1/\left[1+(g\mu_BB_z\tau_s/\hbar)^2\right]$ with $g$ being effective electron $g$-factor and $\mu_B$ being Bohr magneton}. Interestingly, in the structures of crystallographic orientations other than $(110)$, Hanle effect can manifest itself as only partial suppression of spin polarization. Detailed analysis of these effects is beyond the scope of this paper.


Finally we note that in GaAs-based heterostructures, the spin relaxation is usually dominated by the hyperfine interaction~\cite{dyakonov_book,merkulov02}. This makes spin relaxation non-Markovian, or non-monoexponential, so it can not be described by a single time $\tau_s$~\cite{Raikh_2014,shklovskii:193201}. However at moderate magnetic field the spin relaxation is isotropic, which means that the expressions for macroscopic susceptibilities \eqref{sigma_CISP}---\eqref{eq:sigma_SHE} can be applied, where $\tau_s$ should be considered as an ``effective'' or average spin relaxation time.

\section{Conclusion}
\label{sec:concl}

Based on the derived kinetic {equations} describing the coupled spin-charge dynamics, we have identified four regimes of hopping spin transport where SHE, CISP and SGE have different behavior. The numerical simulation shows the map of the spin distribution in the sample in all four regimes. The spin susceptibility is shown to be governed by the ratio of the spin relaxation and hopping times. The percolation analysis being in a very good agreement with the numerical simulations demonstrates how the contributions to the spin effects from each triad in the percolation cluster average over disorder realizations. Application of the perpendicular magnetic field results in damped oscillations of the spin susceptibility where the number of sign changes is also determined by the spin relaxation rate.

\acknowledgments

The  support from  the foundation ``BASIS''  is gratefully acknowledged. The work of D.~S.~S. and L.~E.~G. was supported by 
 Russian Science Foundation (project 17-12-01265).

\appendix



\section{Optimal triads}
\label{App-regimes}

In our percolation analysis we discussed that the contributions of different triads of sites to SHE have an exponentially broad distribution. The effect is dominated by the optimal triads of sites $ikj$ that are defined by optimal geometry of the corresponding triangle $(ikj)$. However this geometry is different in different regimes. {Here} we discuss in details the optimal geometry in all the regimes (A)---(D). 

\begin{figure}
  \centering
  \includegraphics[width=\linewidth]{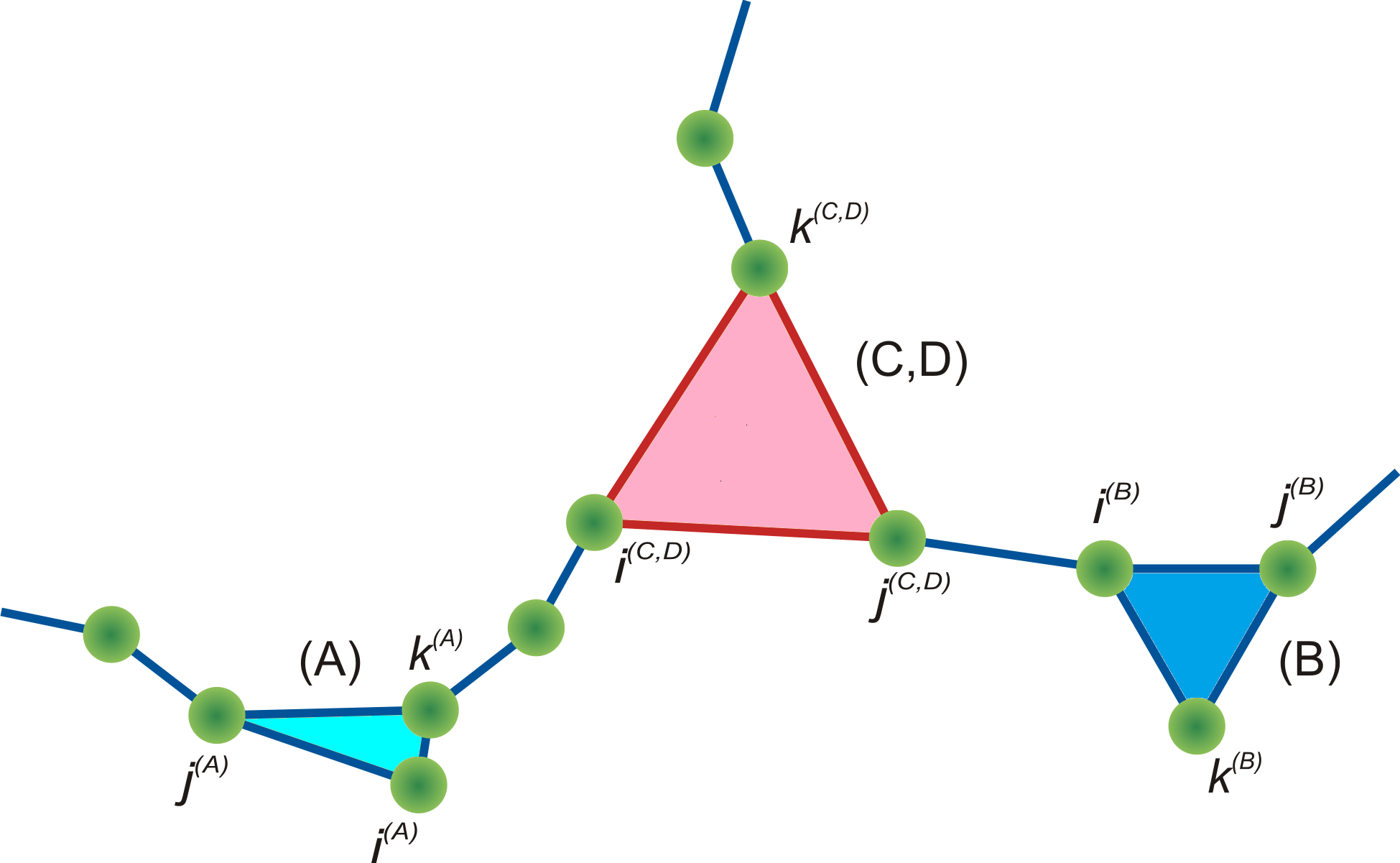}
  \caption{A part of the percolation cluster. The non-critical resistors participating in the cluster are shown with blue color. The critical resistors are shown with red color. The area of the optimal triangles in different regimes is filled in accordance with the {background} colors in Fig.~\ref{fig:f}.  }
  \label{fig:cluster}
\end{figure}

In the regime (A) all the hopping terms can be neglected in Eqs.~\eqref{kin_ikj}. It allows to write the solution explicitly: 
\begin{equation}
\label{Ap1-regA}
\widetilde{S}_i^{(ikj)} = \tau_s' I_{kj} \tau_{kj}\Gamma_{ijk}^{(0)} \propto \\ I_{kj} 
\exp\left( {\frac{r_{kj}-r_{ij}-r_{ik}}{a_b}} \right).
\end{equation}
The similar expressions can be derived for $\widetilde{S}_j^{(ikj)}$ and $\widetilde{S}_k^{(ikj)}$. The exponential part of Eq.~\eqref{Ap1-regA} disappears in the isosceles triangle with
\begin{equation}
  r_{ij} = r_{kj},
  \qquad
  r_{ik}\sim a_b.
  \label{eq:isoscales}
\end{equation}
The long sides $r_{ij}$ and $r_{kj}$ are assumed to {belong to the percolation cluster, see a cyan triangle in} Fig.~\ref{fig:cluster}.
When the geometry of the triangle deviates from the discussed one, $\widetilde{S}_j^{(ikj)}$ exponentially decreases. It is clear from  Eq.~\eqref{Ap1-regA} that it decreases with increasing $r_{ik}$ as $\exp(-r_{ik}/a_b)$. When the triangle $ikj$ deviates from {Eq.~\eqref{eq:isoscales}} the generated spin decreases due to the re-distribution of the currents. Let the side {$r_{kj}$} be larger than $r_{ij}$. The current $I_{kj}$ in this case can be estimated as $I_{kj}=I_{perc}\exp{[-2(r_{kj}-r_{ij})/a_b]}$. It leads to the additional exponentially small term $\exp(-|r_{kj}-r_{ij}|/a_b)$ in the expression for the generated spin. When $r_{ij}>r_{kj}$ the current $I_{kj}$ is equal to $I_{perc}$, but the term  ${\exp(-|r_{kj}-r_{ij}|/a_b)}$ appears in  Eq.~\eqref{Ap1-regA} directly. 

The optimal triangle in the regime (B)  is the equilateral triangle with a side {$r_{opt}$ given by Eq.~\eqref{eq:ropt}, see a blue triangle in} Fig.~\ref{fig:cluster}. The triangle should participate in the percolation cluster. As mentioned in the main text, the discussed geometry is actually the optimal one.  {To prove this} we consider the triangle {with} $y_i=y_j=0$. The side $ij$ of the triangle is {assumed} to be included into the percolation cluster. Its contribution to spin current is  directly related to $\widetilde{S}_k^{(ikj)}$ as ${\cal J}_{ikj}^{(triad)} =\widetilde{S}_k^{(ikj)} y_k/\tau_s$. {We remind that}  in the regime~(B) $\tau_s'\approx \tau_s \ll \tau_d$, and the contribution $\mathcal J_{ikj}^{(med)}$ can be neglected. 

We start with the comparison of the contributions to the spin current of equilateral triangles with different length $r$ of the side. In this case $\tau_{ik}=\tau_{ij}=\tau_{jk}=\tau_0e^{2r/a_b}$, $\tau'_{ikj} = \tau'_0 e^{3r/a_b}$, $\tau_0' = (4/3\hbar)J_0\tau_0^2$. The system of equations~\eqref{kin_ikj}  in this case can be analytically solved:
\begin{equation}\label{Ap1-B1}
\widetilde{S}_k^{(ikj)} = \frac{2}{3} I_{perc} \frac{\tau_s\tau_0}{\tau_0'} \frac{\alpha_{xy} \alpha_{yx} A_{ijk}^z e^{r/a_b}}{e^{2r/a_b} + 3\tau_s/\tau_0}.
\end{equation}
In our analysis we neglect the power law dependence $A_{ijk}^z(r)$ in comparison to exponential dependence $\sim e^{r/a_b}$ of the r.h.s of  Eq.~\eqref{Ap1-B1}. This expression has a maximum at $r = r_{opt}$
\begin{equation}\label{Ap1-B2}
\widetilde{S}_k^{(opt)} = \frac{\alpha_{xy}\alpha_{yx} A_{opt} I_{perc} \tau_0^{3/2}\tau_s^{1/2}}{3\sqrt{3} \tau_0'} ,
\end{equation}
where $A_{opt} = (\sqrt{3}/4)r_{opt}^2$.

Now we should compare a contribution of non-equilateral triangles with expression~(\ref{Ap1-B2}). In this procedure we  consider $r_{ij} = r_{opt}$ and displace the site $k$ from its position corresponding to the equilateral triangle. If we move the site along the $x$ axis, one of the sides $r_{ik}$ and $r_{jk}$ becomes larger than another. Let us consider $r_{ik}<r_{jk}$. In the limit $n_s a_b^2 \ll 1$ it means that $\tau_{ik}\ll\tau_{ij}\ll\tau_{jk}$. However at least for relatively small displacements we can still consider $\tau'_{ikj} = \tau'_0 e^{3 r_{opt}/a_b}$. Also the spin relaxation rate is comparable with $\tau_{ij}$ because $r_{ij}$ is still equal to $r_{opt}$: $\tau_s = \tau_{ij}/3$. In this case we can neglect the term $\tau_s/\tau_{kj}$ in the equation for $S_j$ and disregard {spin diffusion between sites $j$ and $k$}. Also the spin generation at the site $j$ is exponentially smaller than at sites $i$ and $k$ and can be neglected. It leads to direct relation between the polarizations on sites $i$ and $j$: $\widetilde{S}_j^{(ikj)} = \widetilde{S}_i^{(ikj)}/4$. With {Eq.~\eqref{eq:zero}} it allows us to give an explicit expression for $\widetilde{S}_k^{(ikj)}$
\begin{multline}
\widetilde{S}_k^{(ikj)} =  \frac{5}{3} \frac{I_{perc}\tau_s \tau_{ik} \alpha_{xy}\alpha_{yx}A_{opt} \tau_{ik}}{\tau_0' e^{3r_{opt/a}}} \\
 \sim \widetilde{S}_k^{(opt)} e^{2(r_{ik}-r_{ij})/a_b} \ll \widetilde{S}_k^{(opt)}.
\end{multline}

Now we consider the displacement of the site $k$ along the $y$ axis. For this displacement the triangle $ikj$ stays isosceles. Therefore the relation of the spins $\widetilde{S}_i^{(ikj)}$, $\widetilde{S}_j^{(ikj)}$ and $\widetilde{S}_k^{(ikj)}$ is the same as in the case of equilateral triangle $\widetilde{S}_i^{(ikj)}=\widetilde{S}_j^{(ikj)} = -\widetilde{S}_k^{(ikj)}/2$. It leads to the the explicit expression for $\widetilde{S}_k^{(ikj)}$.
\begin{multline}\label{Ap1-B3}
\widetilde{S}_k^{(ikj)} = I_{ij} \tau_{ij} \frac{\alpha_{xy}\alpha_{yx} A_{ikj}^z}{\tau_0'}\frac{\tau_s \tau_{side}}{\tau_{side}+3\tau_s}
\\ \times 
 \exp\left(- \frac{r_{ij} + 2r_{side}}{a_b}\right).
\end{multline}
Here  $r_{ik}=r_{jk} = r_{side}$ and $\tau_{side} = \tau_0\exp(2r_{side}/a_b)$. When $r_{side}=r_{ij}$, the current $I_{ij} = 2I_{perc}/3$, and Eq.~\eqref{Ap1-B3} is reduced to Eq.~\eqref{Ap1-B1}. When $r_{side}$ is larger than $r_{ij}$, the last term in Eq.~\eqref{Ap1-B3} exponentially decreases leading to the exponentially small spin {polarization} $\widetilde{S}_k^{(ikj)}$. When $r_{side} < r_{ij}$  the current $I_{ij}$ becomes small, $I_{ij} \sim I_{perc}\exp{[-2(r_{ij}-r_{side})/a_b]}$, because the resistor $R_{ij}$ is shunted by the resistors $R_{ik}$  and $R_{kj}$. It again leads to the exponentially small spin generation $\widetilde{S}_k^{(ikj)}$. 

The above arguments prove that, in the regime~(B), the dominant contribution to the spin-Hall effect comes from the equilateral triangles with sides $r_{opt}$. $r_{opt}$ increases with $\tau_s$ and becomes larger than $r_c$ at ${\tau_s \gg \tau_0\exp(2r_c/a_b)}$. This spin relaxation time corresponds to the transition from regime (B) to regime (C). In the above analysis we {assumed} that the triangle $ikj$ is included into the percolation cluster. It is not possible when $r_{ij}>r_c$ leading to the upper boundary for the side of the optimal triangle. Therefore in the regimes (C) and (D) the dominant triangles 
have sides $\sim r_c$. 

The spin generation in the regimes (C), (D) is controlled not only by the processes inside the triangle $ikj$ but also by the transition of the spin to the surrounding medium. It leads to the additional restrictions for the position of the triangle $ikj$. All the three sites of the triangle should be parts of the percolation cluster, otherwise the effective transition of spin from the triangle to the medium is impossible. However they should be included in different branches of the cluster, otherwise the resistors of the triangle will be shunted by the non-critical resistors of the cluster. The optimal triangle in regimes (C) and (D) is shown 
in Fig.~\ref{fig:cluster}.  It lies at the intersection of three branches of the percolation cluster. 

In the limit $\tau_s \rightarrow \infty$ our theory of SHE can be mapped on the theory of the ordinary hopping Hall effect. The optimal triangles for the Hall effect are discussed in Ref.~\cite{BryksinReview}. Our predictions for the optimal triangles in regimes (C) and (D) agree with this {work}. 


%

\end{document}